\documentclass[12pt]{iopart}
\usepackage{iopams}
\usepackage{graphics}
\usepackage[dvips]{epsfig}
\usepackage{bm}

\begin{document}

\baselineskip18pt

\newcommand\fat[1]{\bm{#1}}
\newcommand\phidot{\dot{\bm{\phi}}}


\rightline{DAMTP-2004-35}
\rightline{UB-ECM-PF-04-09}
\rightline{hep-th/0404241}

\title{\ \ \ \ \ \ \ \ \ \ \ Cosmology as Geodesic Motion}

\author{Paul K. Townsend\dag\ddag~ and Mattias N.R. Wohlfarth\dag}

\address{\dag\ Department of Applied Mathematics and Theoretical
  Physics,\\Centre for Mathematical Sciences, University of
  Cambridge,\\Wilberforce Road, Cambridge CB3 0WA, U.K.\\[5pt]} 

\address{\ddag\ Instituci\'o Catalana de Recerca i Estudis Avan\c
 cats,\\Departament ECM, Facultat de F\' isica, Universitat de
 Barcelona,\\Diagonal 647, E-08028 Barcelona, Spain} 

\ead{\mailto{P.K.Townsend, M.N.R.Wohlfarth@damtp.cam.ac.uk}}  

\begin{abstract}
For gravity coupled to \( N \) scalar fields, with 
arbitrary potential \( V \),
it is shown that all flat (homogeneous and isotropic) cosmologies correspond
to geodesics in an \( (N+1) \)-dimensional `augmented' target space  of
Lorentzian signature \( (1,N) \), timelike if \( V>0 \), null if \(V=0\) and 
spacelike if \( V<0 \). Accelerating
cosmologies correspond to timelike geodesics that lie within an
`acceleration subcone' of the `lightcone'.
Non-flat (\( k=\pm 1 \)) cosmologies are shown to evolve as projections
of geodesic motion in a space of dimension \( (N+2) \), of 
signature \( (1,N+1) \)
for \( k=-1 \) and signature \( (2,N) \) for \( k=+1 \). This
formalism is illustrated by cosmological solutions of models with an
exponential potential, which are comprehensively analysed; the
late-time behaviour for other potentials of current interest 
is deduced by comparison. 
\end{abstract}
\maketitle



\section{Introduction}

Cosmological models  with scalar field matter have been much studied
in the context of inflation and, more recently, in the context of  the
late-time acceleration that is indicated by  current astronomical
observations (see \cite{PeRa03} for a recent review). 
One theoretical motivation for these studies is that
scalar fields arise naturally from the compactification of
higher-dimensional theories, such as string or M-theory. However, the
type of scalar field potential obtained in these compactifications is
sufficiently restrictive that until recently 
it was considered to be difficult  to get accelerating cosmologies in this
way, although the existence of an accelerating phase in a hyperbolic
($k=-1$) universe obtained by compactification had been  
noted \cite{CC}, and non-perturbative effects in M-theory have since
been shown to allow unstable  de Sitter vacua \cite{KKLT}.  

In an earlier paper, we pointed out that compactification on a compact
hyperbolic manifold with a time-dependent volume modulus naturally
leads to a flat (\(k=0\)) expanding universe that undergoes a
transient period of accelerating expansion \cite{ToWo03}. Numerous
subsequent studies have shown that such cosmological solutions are
typical to all compactifications that involve compact hyperbolic
spaces or non-vanishing \( p \)-form field strengths (flux)
\cite{Oht03a,Roy03,Woh03a,Oht03b,CHNW03,CHNOW03}, and this was
additionally confirmed in a systematic study \cite{Woh03b}.
Furthermore, the transient acceleration in these models is easily
understood  \cite{EmGa03} in terms of the positive scalar field
potential  that both hyperbolic and flux  compactifications produce in
the effective, lower-dimensional, action. This perspective also makes
clear the generic nature of transient acceleration. 

For any realistic application one would want the lower-dimensional
spacetime to be four-dimensional, but for theoretical studies it is
useful to consider a general $d$-dimensional spacetime. Assuming that
we have gravity, described by a metric \(g_{mn}\), coupled to $N$
scalar fields  \( \fat {\phi } \)  taking values in a Riemannian
target space with metric  \( G_{\alpha\beta}\) and with potential
energy \( V \), the effective action must take the form  
\begin{equation}\label{action}
\int  d^dx\, \sqrt{-g}\left( \frac{1}{4}R- \frac{1}{2}g^{mn
}G_{\alpha\beta}(\fat {\phi })\partial _m \phi ^{\alpha}\partial_n\phi
^{\beta}-V(\fat {\phi })\right) 
\end{equation}
where  \( R \) is the (spacetime) Ricci scalar.
We are interested in solutions of the field equations of the action
(\ref{action}) for which the line element has the
Friedmann-Lema\^itre-Robertson-Walker (FLRW) form for a
homogeneous and isotropic spacetime. In standard coordinates, 
\begin{equation}
ds^{2}=-dt^{2}+S(t)^{2}d\Sigma _{k}^{2}\, ,
\end{equation}
where the function \( S(t) \) is the scale factor, and 
\( \Sigma _{k} \) represents
the \( (d-1) \)-dimensional spatial sections of constant curvature $k$.
We normalise $k$ such that it may take values \( k=0,\pm 1 \) for a
Riemann tensor $\bar R_{\bar m\bar n\bar o\bar
  p}=k\left(\bar g_{\bar m\bar o}\bar g_{\bar
  n\bar p}-\bar g_{\bar m\bar p}\bar g_{\bar
  n\bar o}\right)$ with metric $\bar g$ on $\Sigma_k$. 
The scalar fields are taken to depend 
only on time, which is the only choice compatible with the symmetries of FLRW
spacetimes. The universe is expanding if $\partial_t S>0$ and accelerating if 
$\partial_t^2 S>0$. We need only discuss expanding cosmologies 
because contracting cosmologies are obtained by time-reversal.

In some simple cases,
the target space has a one-dimensional factor parametrised by a
dilaton field \(\varphi(\fat{\phi})\), and the potential takes the
form 
\begin{equation}\label{eq.exppot}
V= V_0\, e^{-2a\varphi}
\end{equation}
for some `dilaton coupling' constant $a$ and constant $V_0$. This
model is of special interest, in part because of its  amenability to analysis. 
The special case for which the dilaton is the {\it only} scalar field was
analysed many years ago (for $V_0>0$) 
using the observation that, for an appropriate
choice of time variable, cosmological solutions correspond to
trajectories in the `phase-plane'  parametrised by the first
time-derivatives of the dilaton and the scale factor
\cite{Hal87}. This method  (which has recently been extended to 
potentials that are the sum of two exponentials \cite{JMS04})
allows a simple visualisation of all possible cosmological trajectories. 
Moreover, all  trajectories for flat
cosmologies can be found explicitly \cite{BB88,Tow03} (see also
\cite{Neu03,Vi03,Rus04}), and  a related method allows a visualisation 
of their 
global nature \cite{CLW98,HeWa03}. It was noted in \cite{Hal87} that 
there is both a `critical' and a `hypercritical' value of the dilaton coupling
constant $a$, at which the set of trajectories undergoes a qualitative
change. In spacetime dimension $d$, these values are \cite{BB88, Woh03b}
\begin{equation}\label{exponents}
\alpha_c =\sqrt{\frac{2}{d-2}} \qquad \textrm{and}\qquad
\alpha_h=\sqrt{\frac{2(d-1)}{d-2}}\, .  
\end{equation}
Below the critical value ($a<\alpha_c$) there exists a late-time attractor
universe undergoing accelerating expansion, whereas only transient
acceleration is possible above it.  The hypercritical coupling ($a=\alpha_h$)
separates `steep' exponential potentials
(which arise in flux compactifications) and `gentle' exponential 
potentials (which arise from hyperbolic compactification, in which
case $a>\alpha_c$ so the potential is still too steep to allow
eternal acceleration). 
One aim of this paper is to generalise this type of 
analysis to the multi-scalar case. For $k=0$, this has been
done already for what could be called a `multi-dilaton' model
\cite{Guo}, and for the multi-scalar model with exponential potential 
(\ref{eq.exppot})  \cite{BCGNR03}. Here we consider {\it all} 
cosmological  trajectories (arbitrary $k$) for an exponential
potential of either sign,\footnote{Negative exponential potentials
  are of relevance to pre-big bang and ekpyrotic scenarios. Their
  apparent instability can be overcome by the introduction of higher
  order quantum corrections to the gravitational action, see
  \cite{TBF2002} and references therein.} 
and for any spacetime dimension $d$.
In particular, we find exact solutions for all flat cosmologies
when $V<0$, following the method used in \cite{Tow03} for $V>0$, 
and the exact phase-plane trajectories for all $k$ when 
$V=V_0<0$.

A more ambitious aim of this paper is to determine what can be said
about cosmologies with more general scalar potentials. What kind of
model-independent behaviour can one expect, and  
how generic is the phenomenon of transient acceleration?  
Exponential potentials are simple partly because of the 
power-law attractor cosmologies 
that they permit, but such simple solutions do not occur
for other potentials so other methods are needed. In 
this paper, we develop an alternative method of visualising
cosmological solutions that applies to {\it any} potential, 
and we illustrate it by an application to exponential potentials.  
As we explain, exponential potentials serve as `reference potentials'
in determining the late-time behaviour of cosmologies arising in
a large class of models with other potentials.

Our starting point for the new formalism 
is the observation \cite{GiTo87} that flat cosmological
solutions of gravity coupled to $N$ scalar fields with $V=0$  can be 
viewed as null
geodesics in an `augmented' target space  of dimension $N+1$ with a
metric of Lorentzian signature (see \cite{DHN03} for
related results). Trajectories in this space corresponding to non-flat 
cosmologies are 
neither null nor geodesic. However, we will see that they are 
projections of null 
geodesics in a `doubly-augmented' target space with a metric of 
signature $(1,N+1)$ for $k=-1$ and signature $(2,N)$ for $k=+1$.  

It turns out that the extension of these results to $V\ne0$ is very simple.
Flat cosmologies are again geodesics in the augmented target space but with 
respect to a conformally rescaled  metric; the geodesic is timelike 
if $V>0$ and 
spacelike if $V<0$. Null geodesics are unaffected by the  
conformal rescaling
and hence continue to correspond to ($k=0$) cosmologies for 
$V=0$. The analysis of 
acceleration is also simple in this framework: within the lightcone is an 
`acceleration cone', and if a geodesic enters its acceleration cone then 
the corresponding universe will accelerate. In the case of a flat
target space with an exponential potential, the fixed point geodesics
are just straight lines, and the universe is accelerating if the
straight line lies within the acceleration cone. 
The further extension to $k\ne0$ cosmologies is achieved exactly 
as in the $V=0$ case. Cosmological  trajectories are projections
of geodesics in the doubly-augmented target space with 
respect to a conformally  rescaled metric of signature $(1,N+1)$ 
for $k=-1$ and $(2,N)$ for $k=+1$. The geodesic is timelike if $V>0$, 
null if $V=0$ and spacelike if $V<0$.

The plan of this paper is as follows. Section \ref{sec.mscalar}
reviews, in our conventions, the equations of motion in multi-scalar
cosmologies, and we derive a useful alternative criterion 
for the acceleration of the
scale factor. In section \ref{sec.fixedpoints}, we consider possible
fixed points, concluding that these occur only for exponential
potentials, and we determine the nature of these fixed points 
for arbitrary $d$, $k$ and sign of the potential. 
In section \ref{sec.phase} we discuss the phase-plane 
trajectories and present some exact $V<0$ solutions. In section
\ref{sec.manifold} we develop the interpretation of multi-scalar
cosmologies as geodesic motion. In section \ref{sec.applications}
we introduce the notion of the acceleration cone, and we show how 
the cosmologies arising in exponential-potential models fit into 
the new framework; other potentials of current interest are also considered. 
We summarise in section \ref{sec.discussion}.

\section{Multi-Scalar Cosmologies\label{sec.mscalar}}

We begin by briefly reviewing the equations of motion for 
multi-scalar cosmologies that follow from the action (\ref{action}). 
These equations will be given in two different forms, 
distinguished by the choice of
time coordinate. The second form will  be useful  to our analysis in
the next section of  possible `fixed points', which correspond to power-law 
cosmologies in the first form.

To simplify the equations we define
\begin{equation}
\left(D_t^2\phi\right)^\alpha := \partial_t^2\phi^\alpha +  
\Gamma ^\alpha{}_{\beta\gamma} \partial_t\phi^\beta
\partial_t\phi^\gamma 
\end{equation}
where \( \Gamma^\alpha{}_{\beta\gamma} \) is the Levi-Civit\`a connection
for the target space  metric \(G \), and we introduce  the following 
 `characteristic functions' of the potential V:
\begin{equation}\label{afunctions}
a_{\alpha}(\fat{\phi})=-{1\over2}{ \partial\ln |V|\over
  \partial \phi^\alpha}\,.
\end{equation}
These are the components of a 1-form on the target space dual to a
vector field $\mathbf{a}$. In a vector 
notation, the scalar field equation can now be written as
\begin{equation}\label{eq.sca}
D_t^2{\fat\phi} + \left(d-1\right)H\, 
\partial_t{\fat\phi}  = 2V \mathbf{a}\, , 
 \end{equation}
where $H(t)= \partial_t S/S$ is the Hubble function. The 
Friedmann constraint is
\begin{equation}
S^2\left[|\partial_t{\fat\phi}|^2 + 2V -
  \left(d-1\right)\alpha_c^{-2}H^2\right] =
\left(d-1\right)\alpha_c^{-2}k, 
\end{equation}
where the norm \(|\partial_t{\fat\phi}|\) (and hence 
the inner product) is the one
induced by the target space metric:  
\(|\partial_t{\fat\phi}|^2 =\partial_t{\fat\phi}\cdot 
\partial_t{\fat\phi} =G_{\alpha\beta}\,
\partial_t\phi^\alpha\partial_t\phi^\beta\).  Differentiating the Friedmann
constraint, and using the scalar field equation (\ref{eq.sca}), we deduce the
`acceleration equation'
\begin{equation}
\partial_t^2 S = {2S\over d-1}\left[\alpha_c^2 V -
|\partial_t{\fat\phi}|^2\right],
\end{equation}
where we recall that $\alpha_c$ is the `critical' constant 
given in (\ref{exponents}). Clearly, acceleration 
is possible only if $V>0$. Note too that $H(t)$ cannot vanish unless
either $V<0$ or $k>0$ (or $V=0$, $k=0$ and 
$|\partial_t{\fat\phi}|=0$) so only in these cases can an 
expanding universe recollapse.

For the  following section, it will prove useful to rewrite the above
equations  in terms of a new time coordinate  \( \tau  \), defined as a
function of \( t \) by the relation 
\begin{equation}
d\tau =|V|^{1\over2}dt\,.
\end{equation}
Note that we have allowed for the possibility that  \( V<0 \),
although $V=0$ is excluded. We will also set
\begin{equation}
S(t) = e^{\beta(\tau)}\,. 
\end{equation}
In terms of the variable $\beta$, the condition $\partial_tS>0$ for expansion 
is $\dot\beta>0$, while the condition $\partial_t^2S>0$ for acceleration is
\begin{equation}\label{accel1}
\ddot\beta + \dot\beta^2 - \dot\beta \, \left(\mathbf{a}\cdot  
\dot{\fat\phi}\right) >0\, ,
\end{equation}
where the overdot denotes differentiation with respect to  \(\tau  \). 

In the new time coordinate, the scalar field equation becomes
\begin{equation}\label{newscalar}
D_\tau^2{\fat\phi} =  \left(\mathbf{a}\cdot \dot{\fat\phi}\right) 
\dot{\fat\phi} - \left(d-1\right)\dot\beta\,  \dot{\fat\phi} 
+ 2\, {\rm sign}\, V\, \mathbf{a}\,,
\end{equation}
and the Friedmann constraint becomes
\begin{equation}\label{newFried}
e^{2\beta}|V| \left[| \dot {\fat\phi}|^2 -
\left(d-1\right)\alpha_c^{-2} \dot \beta^2  
+ 2\, {\rm sign}\, V\right] = \left(d-1\right)\alpha_c^{-2} k\,.
\end{equation}
The acceleration equation is now
\begin{equation}\label{accelerationeq}
 \ddot \beta= -{2\over d-1}|\dot{\fat\phi}|^2 - 
\dot\beta^2 + \dot\beta\left(\mathbf{a}\cdot\dot{\fat\phi}\right)
 + {2\alpha_c^2\over d-1}\, {\rm sign}\, V\,, 
 \end{equation}
and the condition (\ref{accel1}) for acceleration is therefore
equivalent to 
\begin{equation}\label{eq.accpot}
|\dot{\fat\phi }|^2 < \alpha_c^2\, \textrm{sign}\, V\, .
\end{equation}
Note that this inequality is independent of \( k \), and hence valid 
for all \( k \). As expected, it can be satisfied only if $V>0$.  

\section{Fixed Points\label{sec.fixedpoints}}

We now wish to determine whether the system of equations 
(\ref{newscalar}, \ref{accelerationeq})
subject to the  constraint (\ref{newFried}), admits any fixed
point solutions for which  
\begin{equation}
\ddot\beta =0\qquad\textrm{and}\qquad D_\tau^2{\fat\phi}= \mathbf{0}\, . 
\end{equation}
We shall see that these conditions are consistent only for exponential
potentials, for which the equations (\ref{newscalar}, 
\ref{accelerationeq}) become those of an autonomous dynamical system,
and we determine the fixed points and their type for all $d$, $k$ and
for either sign of the potential. For this analysis, it is convenient
to introduce the quantity
\begin{equation}\label{omegadef}
\omega = \sqrt{{|\alpha_h^2 -a^2|\over 2}}\,, \qquad a= |\mathbf{a}|\,.  
\end{equation}
Recall that $\alpha_h$ is the `hypercritical' constant given in
(\ref{exponents}).

\subsection{Fixed Point Conditions}

Given $D_\tau^2{\fat\phi}=0$, equation (\ref{newscalar}) yields
\begin{equation}\label{cterms}
\dot{\fat\phi} = c \mathbf{a} \qquad \left(\Rightarrow \ \ 
|\dot{\fat\phi}|^2 = c^2 a^2, \ \  
\mathbf{a}\cdot \dot{\fat\phi} = c\, a^2\right), 
\end{equation}
where $c$ solves the quadratic equation
\begin{equation}\label{quadc}
a^2c^2 -\left(d-1\right) c\dot\beta +2\, \, {\rm sign}\, V =0\,.
\end{equation}
Given $\ddot\beta=0$, equation (\ref{accelerationeq}) yields
\begin{eqnarray}\label{ddotbeta}
2\, {\rm sign}\, V &=& \left(d-2\right)|\dot {\fat\phi}|^2 + 
{1\over2}\left(d-1\right)\left(d-2\right)\left[\dot\beta^2 
- \dot\beta\left(\mathbf{a} \cdot\dot{\fat\phi}\right)\right] \nonumber\\
&=& \left(d-2\right)c^2a^2 + {1\over2}\left(d-1\right)
\left(d-2\right)\left[\dot\beta^2 - ca^2 \dot\beta \right] 
\end{eqnarray}
where we have used (\ref{cterms}) to get to the second line. 
Using this equation to eliminate the $2\,{\rm sign}\, V$ 
term from (\ref{quadc}) we arrive at the quadratic equation
\begin{equation}
\left(\dot\beta - a^2c\right) \left(\dot\beta - \alpha_c^2 c\right)=0\,. 
\end{equation}
There are therefore two types of fixed point:
\begin{itemize}
\item $\dot\beta= \alpha_c^2 c$. Using this in (\ref{quadc}) 
we deduce that $a\ne\alpha_h$ and that
\begin{equation}
c^2 = {2\, {\rm sign}\, V\over \left(\alpha_h^2 -a^2\right)}\,.
\end{equation}
This requires 
\begin{equation}
\!\!\!\!\!\!\!\!\!\!\!\!\! EITHER \ \ \  V>0 \ {\rm and}\  
a<\alpha_h\,, \qquad OR
\ \ \ V<0 \ {\rm and}\ a>\alpha_h\,,
\end{equation}
and 
\begin{equation}
\omega\, \dot{\fat\phi} = \pm  \mathbf{a}\,, \qquad 
\omega\, \dot\beta = \pm \alpha_c^2\,.
\end{equation}
It then follows that
\begin{equation}\label{kzeroFried}
| \dot {\fat\phi}|^2 -\left(d-1\right)\alpha_c^{-2} 
\dot \beta^2  + 2\, {\rm sign}\, V =0\,.
 \end{equation} 
This is just the Friedmann constraint for $k=0$, so 
this type of fixed point occurs only for $k=0$.

\item $\dot\beta= a^2c$. Using this in (\ref{quadc}) we deduce that 
\begin{equation}
c^2 = {\alpha_c^2\over a^2}\ {\rm sign}\,V\,.
\end{equation}
It follows that this type of fixed point can occur only for 
$V>0$, and that
\begin{equation}
\dot{\fat\phi} = \pm \left({\alpha_c\over a}\right)\mathbf{a}\,, 
\qquad \dot\beta = \pm \alpha_c a\,.
\end{equation} 
Using this in the Friedmann constraint we deduce that
\begin{equation}\label{keq}
\alpha_c^{-2}k= e^{2\beta}|V|\left(\alpha_c^2-a^2\right). 
\end{equation}
We may exclude the possibility that $a=\alpha_c$ 
because this yields a $k=0$ fixed point of the type 
already considered, so $k\ne0$ and 
we need $k>0$ for $a<\alpha_c$ and $k<0$ for $a > \alpha_c$. 
Note that $|\dot{\fat\phi}|=\alpha_c$ for this type of 
fixed point, so the fixed point cosmology is neither 
accelerating nor decelerating. 
\end{itemize}

\subsection{Determination of the fixed point type}

The above results are in complete analogy with those of \cite{Hal87,BB88}
for the one-scalar case. To see why, it should first be appreciated that 
the fixed point conditions  have been derived on the assumption that
\(\dot\beta\) is constant, and that \(\dot{\fat\phi}\) is 
covariantly constant. It then follows from 
(\ref{cterms}, \ref{quadc}) that \(\mathbf{a}\) must be 
covariantly constant too. However,  for any space 
that admits a non-zero covariantly constant vector, there 
exist coordinates in which this vector is
{\it constant}, not just covariantly constant. For coordinates 
in which $\mathbf{a}$ is  constant the potential takes the form 
\begin{equation}\label{exppot2}
V= V_0\, e^{-2 \mathbf{a}\cdot {\fat\phi}}\,, 
\end{equation}
which is of the form (\ref{eq.exppot}) with
$a\varphi = \mathbf{a}\cdot {\fat\phi}$. 
Moreover, for any target space on which this potential
is globally defined, one can find new coordinates such that 
\begin{equation}\label{localmetric}
G_{\alpha\beta}\,d\phi^\alpha d\phi^\beta = d\varphi^2 + \widehat
G_{\widehat \alpha \widehat \beta}\left(\chi\right)\, d\chi^{\widehat \alpha}
d\chi^{\widehat \beta}\, , 
\end{equation}
where $\widehat G$ is a metric on a `reduced'  target space with
$(N-1)$ coordinates $\chi$. In these target space
coordinates, the scalar field equations are
\begin{eqnarray}\label{eq.chi}
D_\tau^2\chi &=& a\dot\varphi\dot\chi
-\left(d-1\right)\dot\beta\, \dot \chi\,, \nonumber \\
\ddot\varphi &=& a\dot\varphi^2 - 
\left(d-1\right)\dot\beta\dot\varphi + 2a\, {\rm sign}\, V\,, 
\end{eqnarray}
and the Friedmann constraint is
\begin{equation}\label{friedchi}
\alpha_c^2\left[|\dot\chi|^2 + \dot\varphi^2  + 2\,\textrm{sign}\, 
V_0\right]  -
\left(d-1\right)\dot\beta^2 =
  k\left(d-1\right)e^{2\left(a\varphi-\beta\right)} |V_0|^{-1}\, .  
\end{equation}
These imply the acceleration equation for $\ddot\beta$. If we suppose
that the reduced target space is flat then the scalar field equations,
together with the acceleration equation, define an
autonomous dynamical system:
\begin{eqnarray}\label{autonomous}
\ddot\varphi &=& a\dot\varphi^2 - 
\left(d-1\right)\dot\beta\dot\varphi + 2a\, {\rm sign}\, V\,, \nonumber\\
\ddot\beta &=& -{2\over \left(d-1\right)}\, 
\left[ \dot\varphi^2 + |\dot\chi|^2 - \alpha_c^2\, {\rm sign}\, V\right]
- \dot\beta^2 + a\dot\beta \dot\varphi\,, \nonumber\\
\ddot\chi &=& a\dot\varphi\,  \dot\chi -
\left(d-1\right)\dot\beta\, \dot\chi\,. 
\end{eqnarray}
Note that we may consistently set $\dot\chi=0$ to recover 
the equations of the one-scalar model. As there are no fixed
points with $\dot\chi\ne0$, the fixed points of the multi-scalar 
model are the same as those of the one-scalar model.

If the reduced target space is not flat then a connection term must be added
to the last equation in (\ref{autonomous}). As this introduces 
dependence on $\chi$, the equations no longer define an autonomous dynamical 
system in just $N+1$ variables. It is not
clear how much difference this makes to the results: note that 
the connection term has no effect on the stability of fixed points 
with $\dot\chi=0$ because it is quadratic in $\dot\chi$. However, we
will suppose for the analysis to follow that the reduced target space
{\it is} flat. We may then assume with no essential loss of generality that 
it is also one-dimensional (i.e. $N=2$) so we have an 
autonomous dynamical system for just three variables.  
We linearise about a fixed point with $\dot\varphi=\dot\varphi_0$ and
$\dot\beta=\dot\beta_0$ by writing 
\begin{equation}
\dot\varphi = \dot\varphi_0 + x\,, \qquad \dot\beta = \dot\beta_0 + y\,,
\qquad \dot\chi= z\,.
\end{equation}
The linearised equations for $(x,y,z) =U^T$ take the form $\dot U =
MU$, where $M$ is the $3\times 3$ matrix 
\begin{equation} 
\pmatrix{2a\dot\varphi_0 -
\left(d-1\right)\dot\beta_0 & -\left(d-1\right)\dot\varphi_0 & 0\cr
a\dot\beta_0 -4\left(d-1\right)^{-1}\dot\varphi_0 & 
a\dot\varphi_0 -2\dot\beta_0 & 0 \cr
0 & 0 & a\dot\varphi_0 - \left(d-1\right)\dot\beta_0}. 
\end{equation}
The eigenvalues of this matrix 
determine the nature of the fixed point.
We consider the two types of fixed point in turn:
\begin{itemize}
\item $k=0$. In this case
\begin{equation}
\omega \dot\varphi_0 = \pm a\,, 
\qquad \omega \dot\beta_0 = \pm \alpha_c^2\,.
\end{equation}
The eigenvalues of $M$ are
\begin{equation}
\pm 2\omega^{-1}\left(a^2-\alpha_c^2\right), 
\qquad \pm\omega^{-1}\left(a^2-\alpha_h^2\right), \qquad
 \pm \omega^{-1}\left(a^2-\alpha_h^2\right).
\end{equation}
For an expanding universe we must take the top sign (corresponding, 
for $V>0$, to the upper branch of  the $k=0$ hyperboloid). Then 
we have a stable node (all eigenvalues real
and negative) for $a<\alpha_c$. For $\alpha_c<a<\alpha_h$ we have a
saddle, with one real-negative and two real-positive eigenvalues; the
instability due to the positive eigenvalues is what leads to the
`flatness problem' of standard big-bang cosmology.
 
For $V<0$ we may have $a>\alpha_h$, 
in which case the fixed point is an unstable node (all eigenvalues 
real and positive). All $k=-1$ trajectories start at this unstable
fixed point and end at the stable fixed point on the other branch of
the $k=0$ hyperboloid; as $\dot\beta<0$ at this other fixed point, all 
$k=-1$ universes recollapse to a big crunch singularity. However, for
$a\gg \alpha_h$ there is a long quasi-static period on some of these
trajectories.

\item $k\ne0$. In this case
\begin{equation}\label{fixedpm}
\dot\varphi_0=\pm \alpha_c\,, \qquad \dot \beta_0 = \pm \alpha_c a\,.
\end{equation}
The eigenvalues of $M$ are
\begin{equation}
\!\!\!\pm {\alpha_c\over2}\left[-\left(d-2\right)a + \sigma \sqrt{16 -
\left(d-2\right)\left(10-d\right) a^2}\right], \qquad \mp
2a/\alpha_c\,, 
\end{equation}
where $\sigma=1$ for one eigenvalue and $-1$ for another.   
Take the top sign again. For $a\le \alpha_c$ all
eigenvalues are real but one is positive and the others negative, so we have 
a saddle point. For $a>\alpha_c$ all eigenvalues are real and 
negative provided {\it either} $d\ge10$ {\it or} (if $d\le9$) 
\begin{equation}
a\leq \bar a \equiv 
{4\over\sqrt{ \left(d-2\right)\left(10-d\right)}}\,.
\end{equation}
Otherwise (i.e. $d\le9$ and $a>\bar a$) we have one real negative
eigenvalue and two complex  eigenvalues with negative real parts, 
and hence a stable `spiral-node' (a spiral in the $xy$ plane 
and a node in the $z$ direction). Note that
\begin{equation}
\alpha_c < \bar a \le \alpha_h
\end{equation}
where the equality occurs for $d=9$.
Thus, for any $d$ the $k\ne0$ fixed point is an unstable 
saddle if $a<\alpha_c$, and is stable for $a>\alpha_c$ 
but can be either a node or a spiral-node. For $d\ge10$ it is 
always a node, but for $d=9$ it becomes a spiral-node at 
$a=\alpha_h$ and for $d<9$ it becomes a spiral-node 
at $a=\bar a < \alpha_h$. 

\end{itemize}
In the supergravity context we are restricted to $d\le11$, 
but there is no scalar potential possible in $d=11$ and the 
only possible potential in $d=10$ is a positive exponential (with 
$a> \alpha_h$ so there is no fixed-point solution).

\section{Phase-Space Trajectories}\label{sec.phase}

We now have all the information needed to determine the 
qualitative behaviour 
of all phase-space trajectories.  We begin with the one-scalar model,
obtained by setting $\dot \chi\equiv 0$.  
The trajectories for $V>0$ are sketched in \cite{Hal87} but, for the
reader's convenience, we give a brief summary in words. All $k=0$
trajectories begin, after a big bang singularity, with a period 
of kinetic energy domination with effectively vanishing potential. For 
$a>\alpha_h$ they end up approaching a similar late-time phase,
but for $a<\alpha_h$ they approach the $k=0$ fixed-point solution, 
which is accelerating if $a<\alpha_c$ and decelerating if
$a>\alpha_c$. For $a<\alpha_c$, all $k\ne0$ trajectories 
sufficiently close to
a $k=0$ trajectory also approach the $k=0$ fixed point; in fact, all
$k=-1$ trajectories have this property. The behaviour of a $k=+1$
trajectory is determined by its relation to the separatrix of the
unstable $k=+1$ fixed point (which is the Einstein Static Universe for
$a=0$); those on one side approach the $k=0$ fixed point while those
on the other side represent universes that recollapse to a big
crunch. For $a>\alpha_c$ {\it all} $k=+1$ universes 
recollapse to a big crunch
whereas all $k=-1$ universes now approach a zero-acceleration Milne
universe (the $k=0$ fixed point now being a saddle point). For $a>\bar a$
the $k=-1$ fixed point is an attractor spiral so that all $k=-1$ 
trajectories spiral around a point of zero-acceleration; as observed
in \cite{JMS04}, this leads to eternal oscillation between 
acceleration and deceleration.

For $V<0$ the $k=0$ trajectories were studied in \cite{HeWa03} but
there seems to be no complete study of all trajectories for this case.
We will therefore present some results of the one-scalar model for 
$V<0$ before moving on
to discuss modifications that arise in the multi-scalar case.

\subsection{Some Exact $V<0$ Solutions}

It is well known that anti-de Sitter space is a solution of Einstein's
equations with a negative cosmological constant. This space must
correspond to one of the cosmological trajectories in a model with
$V=V_0<0$ (which implies that $a=0$). Set 
\begin{equation}\label{xydef}
\dot\varphi =x\,, \qquad \dot\beta = y\,,
\end{equation}
to get the equations
\begin{equation}\label{xiequation}
\dot x = -\left(d-1\right)xy\,, \qquad \dot y = -\left[ y^2 + \lambda^2 +
  {2x^2\over d-1}\right]
\end{equation}
where
\begin{equation}
\lambda = {2\over
  \sqrt{\left(d-1\right)\left(d-2\right)} }\,. 
\end{equation}
These equations imply that
\begin{equation}
\left[x^2 - 2\lambda^{-2} y^2 -2\right] = Cx^{2/(d-1)}
\end{equation}
for constant $C$. The curves with $C>0$ are the $k=+1$ trajectories
and those with $C<0$ are the $k=-1$ trajectories. The two 
branches of the $C=0$ curve, a hyperbola, are the $k=0$ trajectories.
The adS solution corresponds to $C=-\infty$, and hence $x=0$; for this
special case we have 
\begin{equation}
\dot y = -\left[y^2 + \lambda^2\right] 
\end{equation}
which has the solution $y= -\lambda\tan\lambda\tau$, and hence
\begin{equation}
e^\beta =S_0 \cos \lambda\tau\,. 
\end{equation}
The Friedmann constraint implies that $k=-1$, as expected, and 
\begin{equation}
|V_0|= {d-1\over 2S_0^2} \qquad \left[\Rightarrow \lambda\tau= 
{\alpha_c t\over |S_0|}\right], 
\end{equation} 
and hence
\begin{equation}
ds^2=-dt^2 + S_0^2 \cos^2\left(\alpha_c t/|S_0|\right)d\Sigma_{-1}^2\,. 
\end{equation}
This is anti-de Sitter space. In contrast to de Sitter space, which
can be realised (for $V=V_0>0$) as a FLRW universe for {\it any} $k$ 
(in particular as the $k=0$ fixed-point solution), anti-de Sitter space
is realised as an FLRW cosmology only for $k=-1$, and it does not
correspond to a fixed point cosmology (at least in the sense of this
paper). 

We now turn to the $k=0$ trajectories for arbitrary $a$. In this case
we have to solve
\begin{equation}
\dot x =-\left(d-1\right) xy + a\left(x^2-2\right)
\end{equation}
subject to the constraint
\begin{equation}
x^2 -2\lambda^{-2} y^2 =2\,.
\end{equation}
The constraint is solved by setting
\begin{equation}\label{xieq}
x = \sqrt{2}\cosh\xi\,, \qquad y = \lambda \sinh \xi
\end{equation}
and the equation of motion becomes
\begin{equation}
\dot\xi = {1\over \sqrt{2}}\left[ \left(a-\alpha_h\right) 
e^\xi - \left(a+\alpha_h\right)e^{-\xi}\right].
\end{equation}
This equation is immediately integrated if $a=\alpha_h$. Having found
$x,y$ we integrate (\ref{xydef}) to deduce that
\begin{equation}
e^{-2\alpha_h\varphi} = |\tau|
  e^{\alpha_h^2\tau^2}\,,\qquad
e^\beta = S_0 \left[|\tau|
e^{-\alpha_h^2\tau^2}\right]^{1\over 2\left(d-1\right)}\,,
\end{equation}
for integration constant $S_0$ (we have set to zero the other
integration constant as it can be absorbed into $V_0$).

If $a\ne\alpha_h$ then we introduce the new variable 
\begin{equation}
\zeta = \sqrt{\left|{\alpha_h-a\over \alpha_h+a}\right|}\ e^\xi\,.
\end{equation}
The two branches of the $k=0$ hyperbola correspond to positive and 
negative $\zeta$; we take $\zeta>0$. We now consider separately 
the cases $a>\alpha_h$ and $a<\alpha_h$:

\begin{itemize}

\item $a<\alpha_h$. In this case the equation for $\zeta$ is
\begin{equation}
\dot\zeta = -\omega \left(\zeta^2 +1\right) \qquad \left[\Rightarrow
\ \zeta= \cot \omega\tau\right], 
\end{equation}
where we recall that $\omega$ is the function of $a$ given in
(\ref{omegadef}). 
Integration of (\ref{xydef}) now leads to 
\begin{eqnarray}
e^{\left(\alpha_h^2-a^2\right)\varphi}  &=& \sin^{\alpha_h+a}
\omega\tau\  \cos^{a-\alpha_h}\omega\tau\,, \nonumber\\
e^\beta &=& S_0 \left[\sin^{\alpha_h+a}
\omega\tau\  
\cos^{\alpha_h-a}\omega\tau\right]^{{\alpha_c^2\over
  2\alpha_h\omega^2}}\,. 
\end{eqnarray}

\item $a> \alpha_h$. In this case the equation for $\zeta$ is 
\begin{equation}
\dot\zeta = \omega\left(\zeta^2-1\right).
\end{equation}
There is a fixed point at $\zeta=1$, so we must distinguish between 
$\zeta>1$ and $\zeta<1$. One finds that
\begin{eqnarray}
\zeta = \cases{ -\coth \omega\tau  & $\zeta>1$
  \cr  -\tanh \omega\tau & $\zeta<1$}\,.
\end{eqnarray}
For $\zeta>1$ integration of (\ref{xydef}) yields
\begin{eqnarray}
e^{\left(\alpha_h^2-a^2\right)\varphi}  &=&
\sinh^{a+\alpha_h}
\omega\tau\  \cosh^{a-\alpha_h}\omega\tau\,, \nonumber\\
e^\beta &=& S_0 \left[  \sinh^{a+\alpha_h}
\omega\tau\  \cosh^{\alpha_h-a}\omega\tau\right]^{\alpha_c^2\over
  2\alpha_h\omega^2}\,. 
\end{eqnarray}
For  $\zeta<1$ integration of (\ref{xydef}) yields
\begin{eqnarray}
e^{\left(\alpha_h^2-a^2\right)\varphi}  &=&
\sinh^{a-\alpha_h}
\omega\tau\  \cosh^{a+\alpha_h}\omega\tau\,, \nonumber\\
e^\beta &=& S_0 \left[  \sinh^{\alpha_h-a}
\omega\tau\  \cosh^{a+\alpha_h}\omega\tau\right]^{\alpha_c^2\over
  2\alpha_h\omega^2}\,. 
\end{eqnarray}

\end{itemize}

\subsection{Multi-scalar Trajectories} 

In the multi-scalar case, the `spectator' fields $\chi$ have an effect
on the scale factor due to the $|\dot\chi|^2$ term in the acceleration
equation, and in the Friedmann constraint (\ref{friedchi}).
Phase space trajectories with non-zero $\dot\chi$ will  depend on the 
reduced target space metric but, as long as there are no new 
fixed points, one may expect the qualitative behaviour to be 
independent of this metric. Let us therefore continue to suppose
that the target space (and hence the reduced target space) is flat.
In this case, all $k=0$ trajectories lie in a hyperboloid in phase
space that separates the $k=-1$ and $k=+1$ trajectories. This is true
for either sign of $V$ but we will now suppose that $V>0$; in this
case, we see from (\ref{eq.accpot}) that acceleration 
occurs whenever the phase-plane trajectory enters the region with 
hyperspherical boundary \( |\dot {\fat\phi }|=\alpha_c\). 
This was observed in \cite{BCGNR03} in the context of an analysis of
$k=0$ trajectories, but it is also valid for non-zero $k$.

The only exponential models with eternally accelerating cosmological
solutions are those for which a fixed point lies inside or 
on the sphere of acceleration, as
happens when $a\le \alpha_c$. For all other values of \( a\), any 
acceleration can be at most transient, and only a subset of the
trajectories undergo even transient acceleration; these are the
trajectories that pass through the sphere of acceleration. The more
scalar fields there are, the more freedom there is to avoid the sphere
of acceleration, so transient acceleration in multi-scalar models is
less generic than it is in the single-scalar model. 
To make this statement quantitative would require an understanding, 
along the lines of \cite{GHS87}, of what the appropriate measure  
might be on the space of trajectories. 

\begin{figure}[htbp]
\bigskip{}
{\par\centering
  \resizebox*{0.95\columnwidth}{!}{\includegraphics{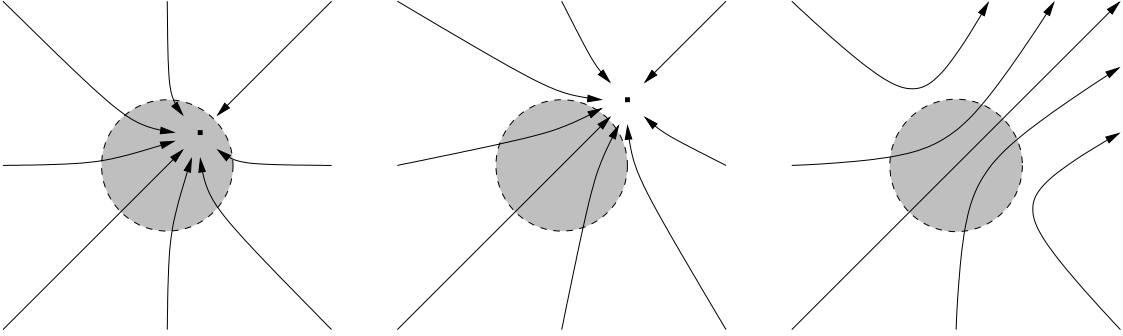}}
  \par} 
\caption{\textit{\label{fig.k0psdia}
These diagrams for \protect\( k=0\protect \) cosmologies  
with two scalar fields moving in a positive, purely 
exponential, potential illustrate the trajectories  in phase 
space. 
The motion takes place in the hyperboloid above the plotted plane, defined by
equation (\ref{kzeroFried}). The constant characteristic functions
\protect\( \mathbf{a}\protect \) of the potential satisfy 
\protect\( a<\alpha _{c}\protect \),
\protect\( \alpha _{c}<a<\alpha _{h}\protect \) 
and \protect\( a>\alpha _{h}\protect \), 
from left to right. The stable direction
\protect\( \mathbf{a}\protect \), corresponding to \(\dot \chi=0\), is 
chosen at an angle \protect\( \pi /4\protect \). 
Trajectories describe accelerating cosmologies when they lie 
inside the sphere
of acceleration marked in grey. }}
\medskip{}
\end{figure}

The most significant fact about a $k=0$ fixed point is whether it lies 
inside or outside the sphere of acceleration. Figure 
\ref{fig.k0psdia} illustrates
a few examples in a model with two scalar fields and a  
positive exponential, scalar potential. 

\section{Augmented Target Space and Geodesic Motion\label{sec.manifold}}

We now develop a method to describe the 
generic evolution of multi-scalar cosmologies. It will be 
convenient to define a new independent variable $\gamma$ by 
\begin{equation}
\left(d-1\right)\beta = \alpha_h\gamma\,, 
\end{equation}
although we will still use $\beta$ where this is more convenient. 
The Friedmann constraint can now be written as
\begin{equation}
\left(\partial_t\gamma\right)^2 - \left|\partial_t{\fat\phi}\right|^2 + 
k\,\alpha_c^{-2}\left(d-1\right)e^{-2\beta} =2V
\end{equation}
and the scalar field equation as
\begin{equation}
D_t^2{\fat\phi} + 
 \alpha_h \left(\partial_t\gamma\right) 
\partial_t {\fat\phi} -2V\mathbf{a} =0\,.
\end{equation}
These two equations imply the acceleration equation
\begin{equation}
\partial_t^2\gamma + {2\over \alpha_h} 
\left|\partial_t{\fat\phi}\right|^2 + 
{\alpha_c\over \sqrt{d-1}} \left[\left(\partial_t\gamma\right)^2 - 
2V\right] =0\,.
\end{equation}
Using the Friedmann constraint to eliminate $2V$ from the scalar 
and acceleration equations we have
\begin{eqnarray}\label{basiceqs}
D_t^2 {\fat\phi} + \alpha_h \left(\partial_t\gamma \right)
\partial_t {\fat\phi} + 
\mathbf{a}\left[ \left|\partial_t{\fat\phi}\right|^2 - 
\left(\partial_t\gamma\right)^2\right] &=&
 k\,\alpha_c^{-2}\left(d-1\right)e^{-2\beta}\,  \mathbf{a}\,, \nonumber\\
\partial_t^2\gamma + \alpha_h \left|\partial_t{\fat\phi}\right|^2 
&=& k\,\alpha_c^{-2}\alpha_h e^{-2\beta}\,. \label{accgamma}
\end{eqnarray}
Before considering the general case (arbitrary $k$ and arbitrary $V$), 
we first discuss the special cases of $V=0$ (but arbitrary $k$) and 
$k=0$ (but arbitrary $V$).

\subsection{$V=0$ Case}

We may consider the $(N+1)$ variables
\begin{equation}
\Phi^\mu  = \left( \gamma, \phi^\alpha\right)
\end{equation}
to be maps from the cosmological trajectory to an  `augmented' target space. 
In this notation, the Friedmann constraint for $V=0$  can be written as
\begin{equation}\label{friedvzero}
G_{\mu\nu}\, \partial_t\Phi^\mu \partial_t\Phi^\nu = k\,\alpha_c^{-2}\left(d-1\right) 
e^{-2\beta}
\end{equation}
where 
\begin{equation} 
G_{\mu\nu}\, d\Phi^\mu d\Phi^\nu =  -d\gamma^2 + G_{\alpha\beta}\, 
d\phi^\alpha d\phi^\beta
\end{equation}
is a {\it Lorentz-signature} metric on the augmented target
space. Using the Friedmann constraint to eliminate the 
$|\partial_t{\fat\phi}|^2$ term in the second equation in
(\ref{accgamma}) yields 
\begin{equation}
\partial_t^2\gamma + \alpha_h \left(\partial_t\gamma\right)^2 =
-k\,\alpha_c^{-2}\left(d-2\right) \alpha_h e^{-2\beta}\,.
\end{equation}
If this is taken together with the scalar field equation then the two
combine into the single (albeit coordinate-dependent) equation 
\begin{equation}
D_t^2\Phi^\mu + \alpha_h\left(\partial_t\gamma\right)\partial_t\Phi^\mu
= k\,\alpha_c^{-2}\left(d-2\right)\alpha_h e^{-2\beta}G^{\mu\nu}\partial_\nu \gamma\,.
\end{equation}
For $k=0$ this is the equation for a geodesic in a non-affine 
parametrisation. In terms of the the new time variable $t'$ 
defined by 
\begin{equation}\label{tprime}
dt' = e^{-\alpha_h\gamma}dt
\end{equation}
we have
\begin{equation}\label{knonzero}
D_{t'}^2\Phi^\mu = k\,\alpha_c^{-2}\left(d-2\right)\alpha_h e^{2\left(d-2\right)\beta}
G^{\mu\nu}\partial_\nu\gamma\,.
\end{equation}
For $k=0$ this is the equation of an affinely parametrised geodesic. 

Although the cosmological trajectory in the augmented target space 
is not a geodesic for $k\ne0$, it can be viewed as the projection 
of a geodesic in a `doubly-augmented' target space of dimension 
$(N+2)$ that is foliated by hypersurfaces isometric to the augmented 
target space. Let 
\begin{equation}
\Psi^A = (\Phi^\mu,\Psi^*)
\end{equation}
be the $(N+2)$ coordinates. We take the metric to be
\begin{equation}\label{doubleaug}
G_{AB}\, d\Psi^Ad\Psi^B= G_{\mu\nu}\, d\Phi^\mu d\Phi^\nu  + 
G_{**}\left(d\Psi^*\right)^2
\end{equation}
where
\begin{equation}\label{gstarstar}
G_{**} = -k\,\alpha_c^{-2} e^{-2\left(d-2\right)\beta}\,.
\end{equation}
As we now have one more variable we also need another equation. 
We take this to be the `projection equation'
\begin{equation}\label{phistar}
\partial_{t'}\Psi^* = \left(d-1\right)^{1\over2}\, 
e^{2\left(d-2\right)\beta}\,. 
\end{equation}
With this choice, the Friedmann constraint becomes
\begin{equation}
G_{AB}\, \partial_{t'}\Psi^A \partial_{t'}\Psi^B =0\,, 
\end{equation}
and the combined scalar field and acceleration equations 
(\ref{knonzero}) are equivalent to the single geodesic equation
\begin{equation}
D_{t'}^2\Psi^A =0\,. 
\end{equation}
Thus, cosmological trajectories for $V=0$ are null geodesics 
in the doubly-augmented target space.
Note that the signature of this space is Lorentzian if $k<0$ 
but non-Lorentzian (with two `times') if $k>0$.
Because $\Psi^*$ is not constant, the motion is not 
restricted to a single hypersurface of constant $\Psi^*$
and this accounts for the fact that the projection of the 
motion onto one such hypersurface is not geodesic. This argument fails
for $k=0$, for which the projected motion {\it is} geodesic, because
the metric on the doubly-augmented target space is degenerate when
$k=0$.

\subsection{$k=0$ Case}

The Friedmann constraint for $k=0$, but arbitrary $V$, can be written as
\begin{equation}
G_{\mu\nu} \partial_t\Phi^\mu \partial_t\Phi^\nu = -2V\,.
\end{equation}
Thus, the cosmological trajectory in this space is timelike for $V>0$, 
null for $V=0$ and spacelike for $V<0$. As we have seen, it is a null 
{\it geodesic} when $V=0$. When $V\ne0$ the cosmological trajectories 
are no longer geodesic with respect to the metric $G_{\mu\nu}$, but 
they {\it are} geodesics with respect to the conformally-rescaled metric
\begin{equation}
\widetilde G_{\mu\nu}= \Omega^2 G_{\mu\nu}
\end{equation}
where the conformal factor is
\begin{equation}\label{conformalfac}
\Omega=\cases{{1\over\sqrt{2}}\sqrt{\left|V\right|} \, 
e^{\alpha_h \gamma} & $V\ne0$ \cr 
{1\over\sqrt{2}}\ \, e^{\alpha_h \gamma} & $V=0$\,.}
\end{equation}
In fact, one finds that the equations (\ref{basiceqs}) are equivalent, 
for $k=0$, to the equation
\begin{equation}\label{autoparallel}
\widetilde D_t^2 \Phi^\mu \equiv 
\partial_t^2\Phi^\mu + \widetilde \Gamma^\mu{}_{\nu\rho} 
\partial_t\Phi^\nu\partial_t\Phi^\rho
 = f \partial_t\Phi^\mu\,,
\end{equation}
where $\widetilde\Gamma^\mu{}_{\nu\rho}$ is the Levi-Civit\`a connection 
for $\widetilde G_{\mu\nu}$, and\footnote{It should be understood here that $\mathbf{a}={\bf 0}$ 
when $V=0$ as $\ln V$ is undefined in this case.} 
\begin{equation}
 f= \alpha_h \partial_t\gamma -2\mathbf{a}\cdot \partial_t{\fat\phi}\,. 
\end{equation}
This is the equation 
for a geodesic in a non-affine parametrisation; noting that 
\begin{equation}
f = \partial_t \Phi^\mu \, \partial_\mu \ln \left(2\Omega^2
e^{-\alpha_h\gamma}\right) 
 \end{equation}
we deduce that the geodesics are affinely parametrised by a new 
time-coordinate $\hat t$ for which
\begin{equation}
d\hat t =  \left(2\Omega^2 e^{-\alpha_h\gamma}\right)\, dt\,. 
\end{equation} 
In other words,
\begin{equation}\label{affinegeo}
\widetilde D_{\hat t}^2 \Phi^\mu =0\,.
\end{equation}
The Friedmann constraint now takes the form
\begin{equation}
\widetilde G_{\mu\nu} \partial_{\hat t}\Phi^\mu \partial_{\hat t}\Phi^\nu = 
\cases{ -\, {\rm sign}\, V & $V\ne0$ \cr 0 & $V=0$\,.}
\end{equation}
Notice that $\hat t$ differs from $t'$ as defined (for $V=0$) 
by (\ref{tprime}). In fact,
\begin{equation}\label{ttprime}
d\hat t = 2\Omega^2 dt'\,. 
\end{equation}
This difference occurs  because the affine parameter of a null 
geodesic is affected by a conformal rescaling of the metric, and 
the metric we are now considering is the conformally rescaled one 
$\widetilde G$; clearly a null curve that is geodesic with respect to 
$G$ is also geodesic with respect to any other conformally 
equivalent metric, such as $\widetilde G$, but the affine parametrisation 
will differ in general. In contrast, when $V\ne0$ the cosmological 
trajectories in the augmented target space are geodesics  with 
respect to a unique metric (up to homothety) in the class of 
metrics that are conformally equivalent to $G$, and 
this metric is $\widetilde G$.

\subsection{General Case: Arbitrary $k$ and $V$.}

The steps that led to (\ref{affinegeo}) for $k=0$ lead, for general  $k$, 
to the equation
\begin{equation}\label{general}
\widetilde D_{\hat t}^2 \Phi^\mu = \cases{- {1\over4}k\,\alpha_c^{-2}\left(d-1\right) 
e^{-2\beta} |V|^{-1} \widetilde G^{\mu\nu} \partial_\nu 
\left(\ln e^{2\beta}|V|\right) & $V\ne0$ \cr 
- {1\over4}k\,\alpha_c^{-2}\left(d-1\right)e^{-2\beta} \widetilde G^{\mu\nu}
\partial_\nu \left(\ln e^{2\beta}\right) & $V=0$\,.}
\end{equation}
Although this is not the equation of a geodesic when $k\ne0$, 
it is the projection of a geodesic in the doubly-augmented target 
space with respect to the conformally rescaled metric
\begin{equation}
\widetilde G_{AB} = \Omega^2 G_{AB}
\end{equation}
where the conformal factor $\Omega$ is as given in
(\ref{conformalfac}), 
and $G_{AB}$ is as given in (\ref{doubleaug}). From (\ref{phistar}) 
and (\ref{ttprime}) we deduce that
\begin{equation}\label{phistarhat}
\partial_{\hat t}\Psi^* = 
\cases{ \left(d-1\right)^{1\over2}e^{-2\beta} \, |V|^{-1} & $V\ne0$ \cr
\left(d-1\right)^{1\over2} e^{-2\beta} & $V=0$\,.}
\end{equation}
Given this, one then finds that (\ref{general}) is equivalent to 
the geodesic equation
\begin{equation}
\widetilde D_{\hat t}^2\Psi^A =0
\end{equation}
and that the Friedmann constraint is 
\begin{equation}
\widetilde G_{AB}\, \partial_{\hat t}\Psi^A\partial_{\hat t}\Psi^B = 
\cases{ -{\rm sign}\, V & $V\ne0$ \cr
0 & $V=0$\,.}
\end{equation}
Thus, the general cosmological trajectory is a geodesic in the 
doubly-augmented target space, one that is timelike for $V>0$, 
null for $V=0$ and spacelike for $V<0$.

\section{Applications}\label{sec.applications}
We now consider some applications of the formalism just developed. 
As we are particularly interested in accelerating cosmologies we 
first consider the implications of the acceleration condition. We will 
then see how to interpret the fixed point cosmologies that arise for 
exponential potentials. This turns out to be useful when considering 
the asymptotic behaviour of more general potentials. 

\subsection{The Acceleration Cone}

We first show how the condition for acceleration 
(\ref{eq.accpot}) acquires a geometrical meaning in the new framework.   
We will now assume that $V>0$, as acceleration can not otherwise
occur, 
and before proceeding we note that the Friedmann constraint (\ref{newFried})
can be written as 
\begin{equation}\label{friedform}
\dot\gamma^2 = |\dot{\fat\phi}|^2  + 2  -
k\,\alpha_c^{-2}\left(d-1\right) V^{-1} e^{-2\beta}\,. 
\end{equation}

We now introduce the following `acceleration' metric on the augmented 
target space,
\begin{equation}
G^{\rm acc}_{\mu\nu}\, d\Phi^\mu d\Phi^\nu = -{1\over d-1} \, 
d\gamma^2 + G_{\alpha\beta}\, d\phi^\alpha d\phi^\beta\,,
\end{equation}
and the corresponding conformally rescaled metric
\begin{equation}
\widetilde G^{\rm acc}_{\mu\nu} = \Omega^2 G^{\rm acc}_{\mu\nu}, 
\end{equation}
where the conformal factor is the same as before. Noting that
\begin{equation}
d\hat t = \sqrt{2}\, \Omega\, d\tau
\end{equation}
one finds that
\begin{equation}
\widetilde G^{\rm acc}_{\mu\nu}\, \partial_{\hat t}\Phi^\mu \partial_{\hat
  t}\Phi^\nu  
={1\over2}G^{\rm acc}_{\mu\nu}\, \dot \Phi^\mu \dot\Phi^\nu =  
{1\over2}|\dot{\fat\phi}|^2 -{1\over 2\left(d-1\right)}\, \dot\gamma^2\,.
\end{equation}
Using the Friedmann constraint in the form (\ref{friedform}) 
to eliminate the $\dot\gamma^2$ term, we deduce that
\begin{equation}
\widetilde G^{\rm acc}_{\mu\nu}\, \partial_{\hat t}\Phi^\mu \partial_{\hat
  t}\Phi^\nu  
= {1\over \alpha_h^2}\left[ |\dot{\fat\phi}|^2 -\alpha_c^2\right] + 
{k\over 2\alpha_c^{2}V}\, e^{-2\beta}\,.
\end{equation}
Recalling that acceleration occurs when $|\dot{\fat\phi}|^2
<\alpha_c^2$, 
we see that  the condition for acceleration is equivalent, for $k=0$, to
\begin{equation}
\widetilde G^{\rm acc}_{\mu\nu}\, 
\partial_{\hat t}\Phi^\mu \partial_{\hat t}\Phi^\nu  < 0\,. 
\end{equation}
Geometrically, this states that a universe corresponding to a timelike 
geodesic trajectory is accelerating when its tangent vector  lies 
within a subcone of the lightcone defined by the acceleration 
metric on the augmented target space. 

As one might expect, an analogous result holds for $k\ne0$ in terms of 
the doubly-augmented target space. In this case, the acceleration metric is
\begin{eqnarray}
\!\!\!\!\!\!\!\!\!\!\!\!G_{AB}^{\rm acc}\, d\Psi^A d\Psi^B &=&
G_{\mu\nu}^{\rm acc}\,  
d\Phi^\mu d\Phi^\nu + {1\over \left(d-1\right)} 
G_{**}\left(d\Psi^*\right)^2\nonumber\\
&=& G_{\alpha\beta}\, d\phi^\alpha d\phi^\beta + 
{1\over \left(d-1\right)} \left[ - d\gamma^2  -
k \,\alpha_c^{-2}e^{-2\left(d-2\right)\beta}\left(d\Psi^*\right)^2\right]. 
\end{eqnarray}
Using the projection equation (\ref{phistarhat}) one finds that 
\begin{equation}
\widetilde G_{AB}^{acc}\, \partial_{\hat t}\Psi^A \partial_{\hat t}\Psi^B 
=  {1\over \alpha_h^2}\left[ |\dot{\fat\phi}|^2 -\alpha_c^2\right] , 
\end{equation}
where $\widetilde G^{\rm acc}_{AB} = \Omega^2 G^{\rm acc}_{AB}$ is the
conformally rescaled acceleration metric. The condition for acceleration is
therefore equivalent to
\begin{equation}
\widetilde G_{AB}^{\rm acc}\, \partial_{\hat t}\Psi^A \partial_{\hat
  t}\Psi^B <0\,.  
\end{equation}
This states that a universe is accelerating when the tangent to 
its geodesic trajectory in the doubly-augmented target space lies 
with an acceleration subcone of the lightcone.

\subsection{Trajectories for Exponential Potentials\label{sec.costraj} }

We now discuss how the cosmologies arising for an exponential 
potential of the form (\ref{exppot2})
fit into the new geometrical framework. For reasons given earlier  
we assume in this case that the target space is flat, in which 
case the augmented target space metric $G$ is also flat, and hence 
$\widetilde G$ is conformally flat, with conformal factor
\begin{equation}
\Omega \propto e^{\alpha_h\gamma -\mathbf{a}\cdot{\fat\phi}} 
\qquad\left[\Rightarrow \ 
\partial_\mu \Omega \propto \left(\alpha_h, -\mathbf{a}\right)\right]. 
\end{equation}

Our first task is to determine the trajectories in the augmented
target space that are associated with the fixed point solutions. 
Consider first the $k=0$ fixed point. This solution 
has the property that the linear function 
\begin{equation}
J(\Phi)= a^2\gamma -\alpha_h \, \mathbf{a}\cdot {\fat \phi}  
\end{equation}
is time-independent: $0=\dot J = \dot\Phi^\mu\partial_\mu J$. In other
words, $\dot \Phi$ is a vector tangent to the hyperplanes of constant 
$J$. For $N=1$ there is only one such direction, which is determined
by the gradient of $\Omega$. Thus, 
$\dot \Phi^\mu \propto G^{\mu\nu}\partial_\nu \Omega \propto
\left(\alpha_h, a\right)$ for $N=1$. 
In fact, this remains true for arbitrary $N$ as a direct calculation shows: 
\begin{equation}\label{eq.fpdirec}
\dot \Phi^\mu = \pm \omega^{-1}\,
\left(\alpha_h, a^\alpha \right).
\end{equation}
A  fixed point solution for $k\ne0$ 
has the property that the linear function
\begin{equation}
K(\Phi) = \alpha_c^2 \gamma -\alpha_h\, \mathbf{a}\cdot {\fat \phi}
\end{equation}
is time-independent. As surfaces of constant $K$ are 
hyperplanes, a $k\ne0$ fixed point trajectory is {\it also} a
straight line, as a direct computation confirms:
\begin{equation}\label{direction.nonzerok}
\dot\Phi^\mu= \pm \left(\sqrt{\left(d-1\right)}  
\, a, \ {\alpha_c\over a}\, \mathbf{a}\right). 
\end{equation}
Thus, fixed point solutions are (particular) straight-line trajectories 
in the augmented target space. In addition, as follows from our
earlier results, the $k=0$ straight-line trajectories are geodesics
with respect to the metric $\widetilde G= \Omega^2 G$, whereas this is not true
for $k\ne0$. This difference can be understood as
follows: as we have assumed a flat target space, the metric $G$ 
is flat and {\it all} straight lines are geodesics with respect 
to it. However, the relevant metric is $\widetilde G$, and the only
geodesic of $G$ that is also a geodesic of $\widetilde G$ is the line
of steepest descent of the function $\Omega$; as we have seen, this
is precisely the direction of the $k=0$ fixed point trajectory.

Note that the line determined by (\ref{direction.nonzerok}) is a
generator of the acceleration cone (as expected from the
zero-acceleration property of the $k\ne0$ fixed point). 
More generally, if a straight line trajectory lies within the 
acceleration cone then it corresponds to an eternally accelerating 
universe, and if it lies outside 
the acceleration cone ($a>\alpha_c$) then it corresponds to an eternally
decelerating universe.  
In the latter case it might also lie outside the `lightcone'
($a>\alpha_h$), in which case the geodesic corresponds to the $k=0$ 
fixed point for $V<0$.  
Note that a change in the value of the constant vector $\mathbf{a}$
effects a Lorentz transformation of the $k=0$ straight-line trajectory
in the augmented target space. Such a transformation can take 
any timelike line into any other  timelike line, and any 
spacelike line into any other spacelike line, but it cannot take 
a spacelike line to a timelike one or vice versa 
(as expected from the fact that
this requires a change of sign of the potential).
Also, it cannot take a timelike or spacelike 
line into a null line, a fact that is consistent with the 
absence of a fixed-point solution for $a=\alpha_h$.

\begin{figure}[htbp]
{\par\centering \resizebox*{!}{0.3\textheight}
{\includegraphics{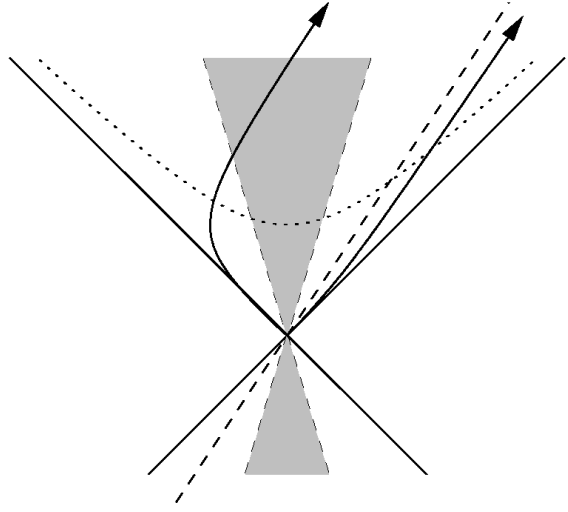}} \par}
\caption{\textit{\label{fig.k0geodesics}
Generic evolution of the single-scalar, \protect\( k=0\protect \), 
cosmologies as geodesics on the augmented target space. The plot shows
the \protect\( \{\gamma,\varphi\}\protect \)-plane in the case of a positive
exponential potential \protect\( V\protect \) with constant characteristic
functions satisfying \protect\( \alpha _{c}<a<\alpha _{h}\protect \).
The decelerating fixed point solution is shown as the dashed straight line
outside the acceleration cone. Both generic geodesics start close to a null
direction when the kinetic energy dominates and are attracted to the 
fixed point
direction at late times. The geodesic on the left  
goes through a regime of transient
acceleration when its tangent vector is steep enough to 
lie within the acceleration cone.}} 
\end{figure}

Generic geodesics, corresponding to generic \(k=0\) cosmologies,
are not straight lines in the extended target space, but they still
have a simple description. The regime where the solutions are
dominated by kinetic energy is given by geodesics with tangent 
vectors having large
\(|\partial_{\hat t}{\fat\phi}|\).  This means that the generic 
geodesics start out in null 
directions, as could be anticipated from the fact that \(k=0\) 
cosmologies are null 
geodesics when \(V=0\). Let us now follow the subsequent 
evolution for the \(V>0\) case. 
As the potential becomes more important, the geodesics bend into  
the timelike cone, ultimately approaching the timelike straight line
corresponding to the fixed point solution if
\(a<\alpha_h\). On the other hand, if
\(a\geq\alpha_h\), then the geodesic  ultimately 
approaches a null 
straight-line geodesic. For the single-scalar case, this 
behaviour is shown in figure 
\ref{fig.k0geodesics}. Cosmologies accelerate precisely 
when their corresponding 
geodesics bend their tangent vectors into the acceleration cone.   

For $k\ne0$, cosmological trajectories are no longer geodesics in the
augmented target space but they are projections of geodesics 
in the doubly-augmented target space. 
The vector tangent to the geodesic corresponding to a $k\ne0$ fixed 
point is 
\begin{equation}\label{fixedcurve}
\partial_{\hat t} \Psi^A = \pm 
{1 \over \sqrt{2}\Omega}\left(\sqrt{\left(d-1\right)}  
\, a, \ {\alpha_c\over a}\, \mathbf{a}, \ \pm\sqrt{\left(d-1\right)}
\, V_0^{-{1\over2}}e^{\mathbf{a}\cdot {\fat\phi} +
  (d-3)\beta}\right) 
\end{equation}
where we have used (\ref{phistarhat}) to get the last entry.
Note that this is {\it not} a constant $(N+2)$-vector (even allowing
for the fact that $a\varphi-\beta$ is constant at the 
fixed point) so the fixed
point solution is not a straight line in the doubly-augmented target
space, as was to be expected because its metric is neither flat nor
conformally flat. However, using (\ref{fixedcurve}), and then
(\ref{keq}), one finds that 
\begin{equation}
\widetilde G_{AB}^{\rm acc}\,  
\partial_{\hat t}\Psi^A \partial_{\hat t}\Psi^B = 
{1\over2}\left[ \alpha_c^2 -a^2 -k\,\alpha_c^{-2}|V|^{-1}e^{-2\beta}\right] =0\,. 
\end{equation}
This again confirms the zero acceleration of the $k\ne0$ 
fixed point cosmologies.  

\subsection{Asymptotic Behaviour for Generic Potentials}

The asymptotic (late-time) behaviour of a large class of cosmologies 
in models with rather general potentials can be determined by 
comparison with the asymptotic  behaviour of cosmologies in models
with exponential potentials. To demonstrate this, we discuss the case 
of flat cosmologies with positive potentials. In this case, 
the scalar fields must approach
\(\fat{\phi}\rightarrow\fat{\phi}_0\) at late times, where some
components of the constant $N$-vector \(\fat{\phi}_0\) may be infinite. 
The characteristic functions of the potential in this limit 
are also constant: 
\begin{equation}
\mathbf{a}_{0} = \lim_{\fat{\phi} \rightarrow
\fat{\phi}_0}\mathbf{a}(\fat{\phi})\,.  
\end{equation}
The absolute value \(a_0=|\mathbf{a}_0|\) determines the late-time
behaviour, by comparison with the critical and hypercritical 
exponents: if \(a_0<\alpha_c\), then there will be late-time
acceleration, with the trajectory approaching an approximate
accelerating attractor; obvious examples are models for which $V$ has a
strictly positive lower bound, so most models of interest will
be those for which $V$ either tends to zero or has a minimum at zero.  
For \(\alpha_c<a_0<\alpha_h\), there is late-time deceleration
corresponding to the existence of an approximate late-time
decelerating attractor. The asymptotic geodesics
in both these cases are timelike. In contrast, for \(a_0>\alpha_h\) 
the timelike geodesic asymptotes a null (and hence decelerating) 
trajectory.
  
To illustrate the above observations, we consider the following single-scalar
examples. 
\begin{itemize}

\item Inverse power law potentials:
\begin{equation}
V=1/\varphi^n\,.
\end{equation}
These potentials have been much studied in quintessence models (see
e.g. \cite{PeRa03}). They
also arise from dynamical breaking of supersymmetry \cite{ADS84}. 
With \(\varphi_0=\infty\), one finds \(a_0=0\). This implies late-time
acceleration, {\it even though the potential tends to zero at late-times}.

\item That any other value of \(a_0\) may occur is illustrated by the 
potential
\begin{equation}
V=\varphi^{2m}e^{-2a\varphi}\,,
\end{equation}
for which \(a_0=a\) if we suppose that $\varphi\rightarrow\infty$
at late times. For such cosmologies the asymptotic behaviour
is the same as it would be in the exponential potential model with
dilaton coupling constant $a$. 

\item 
It may happen that at late times the scalar fields become trapped near
a minimum of the potential. Near such a minimum, which we may assume
to occur at the origin of field space, the potential takes the form
\begin{equation}
V=\varphi^{2n}+\lambda
\end{equation} 
for integer \(n>0\). For positive \(\lambda\) one finds
\begin{equation}
a=-\frac{n\phi^{2n-1}}{\varphi^{2n}+\lambda}
\end{equation}
which tends to zero (i.e. $a_0=0$) as \(\varphi\rightarrow 0\) for any
\(\lambda>0\). This implies late-time acceleration, due to
the effective cosmological constant \(\lambda\). However, we may now
consider the limit $\lambda\rightarrow 0$, in which case 
\(a_0=\infty>\alpha_h\), implying that the late-time trajectory is
null, and hence decelerating. 

\end{itemize}

\section{Discussion\label{sec.discussion}}

In this paper we have developed a geometrical method for the 
classification and visualisation of homogeneous isotropic cosmologies 
in multi-scalar models with an arbitrary scalar
potential $V$. The method involves two steps. In the first
step, the target space parametrised 
by the $N$ scalar fields is augmented to a larger $(N+1)$-dimensional 
space of Lorentzian signature, the logarithm of the scale factor 
playing the role of time. This is reminiscent of the role of the 
scale factor in mini-superspace models of quantum cosmology, 
especially if one views the scalar fields of our model as 
moduli-fields for extra dimensions. In this case, the 
Lorentzian metric on the augmented target space is the one 
induced from the Wheeler-DeWitt metric on the space of 
higher-dimensional metrics. This was the point of view adopted in 
\cite{GiTo87}, where it was also observed that when $V=0$ 
all flat ($k=0$) cosmologies are null geodesics in this 
metric (an observation that goes back to work of DeWitt 
on Kasner metrics \cite{DW67}). When $V\ne0$, 
flat cosmologies again correspond to trajectories in the 
augmented target space but they are neither null nor geodesic. 
However, as we have shown, {\it all} flat cosmologies are 
geodesics with respect to a {\it conformally rescaled} metric 
on the augmented target space, the 
conformal factor depending both on the scale factor and the 
potential. The conformal rescaling has no effect on null 
geodesics, of course, so flat cosmologies are still null 
geodesics whenever $V=0$,  but now the trajectories of flat 
cosmologies are timelike  geodesics when $V>0$ and spacelike 
geodesics when $V<0$. This is true not only for a potential 
of fixed sign but also for one that changes sign; 
in this case the tangent to the geodesic becomes 
null just as the scalar fields take values for which 
$V=0$.

The second step, which is needed for non-flat ($k=\pm1$) 
cosmologies, is to further enlarge the target space to a 
`doubly-augmented' target space foliated by copies of the 
augmented target space. One can choose the metric on this 
space, of signature $(1,N+1)$ for $k=-1$ and $(2,N)$ for 
$k=1$ and degenerate for $k=0$, such that geodesics yield 
(on projection) 
cosmological trajectories for {\it any} $k$ when projected 
onto a given hypersurface, and the geodesics are again timelike, 
null or spacelike according to whether $V$ is positive, 
zero or negative. This general construction thus includes 
all the previous ones as special cases.

Accelerating cosmologies have a simple interpretation in 
this new framework. Within the lightcone (defined by the metric with respect
to which the trajectories are geodesics) there is an acceleration 
subcone. A given trajectory corresponds to an accelerating 
universe whenever its tangent vector lies within the acceleration
cone. In particular, a flat cosmology undergoes acceleration 
whenever its geodesic trajectory bends into the acceleration 
cone within the lightcone of the augmented target space. This 
can happen only if the trajectory is timelike, which it will 
be if the potential is positive. 

We have also presented in this paper a complete treatment 
(complementing many previous studies) of cosmological trajectories, 
for any $k$ and any spacetime dimension $d$, for the special case 
of simple exponential potentials. These potentials are 
characterised by a sign (the potential may be positive or 
negative) and a (dilaton) coupling constant $a$ (the magnitude 
of an $N$-vector  coupling constant $\mathbf{a}$). As has long 
been appreciated \cite{Hal87}, cosmological solutions in such 
models correspond, for an appropriate choice of time parameter, 
to trajectories of an autonomous dynamical system  in a 
`phase space' parametrised by  the time-derivatives of the 
scale factor and scalar fields. The qualitative features of 
these trajectories (which should not be confused with 
trajectories in the `augmented target space' parametrized 
by the fields themselves)  are determined by the position 
and nature of any fixed points, which are of two types. 
There is always a $k=0$ fixed point unless $a=\alpha_h$ 
(the `hypercritical' value of $a$ defined by this property) 
but it occurs only for $V>0$ if $a>\alpha_h$ and only for 
$V<0$ if $a>\alpha_h$. There is also a $k\ne0$ fixed point 
if $V>0$, which coincides with the $k=0$ fixed point when 
$a=\alpha_c$, where $\alpha_c$ is the `critical' value of 
$a$ (defined as the value at which fixed point cosmologies have 
zero acceleration). 

These fixed point cosmologies have a very 
simple interpretation as trajectories in the augmented target 
space: they are straight lines. In the $k=0$ case these lines are also
geodesics. For $V<0$ these geodesic lines lie outside the lightcone.
For $V>0$ they lie inside the lightcone, but may lie either
inside or outside the acceleration cone. For models obtained by (classical)
compactification from a higher dimensional theory without a scalar
field potential, the $k=0$ geodesic line always lies outside the 
acceleration cone, so only transient acceleration is possible in
these models. For a single scalar field the transient  
acceleration is generic in the sense
that it is a feature of all $k=0$ trajectories when $a>\alpha_h$ 
and of some when $a<\alpha_h$.  In contrast, in the multi-scalar 
case, there are trajectories that correspond to eternally 
decelerating universes even when $a>\alpha_h$, so transient 
acceleration is less generic for more than one scalar field.

Although exponential potentials are of limited phenomenological
value, they are important as `reference potentials' in 
determining the asymptotic behaviour of
cosmologies in models with other potentials, such as inverse power
potentials. Essentially, any potential that falls to zero slower than
the critical exponential potential (as do inverse power potentials)
will lead to late-time eternal acceleration. Any potential that falls
to zero faster than the hypercritical exponential potential will have
a late-time behaviour that is approximately that of a model with zero
potential. 

Of course, there is no evidence for the existence of cosmological
scalar fields, but there {\it is} evidence that the expansion of the
universe is accelerating and hence for dark energy. Whatever produces
this energy, it seems reasonable to suppose that it can be modeled by
scalar fields. If so, it is possible that future observations may 
be interpreted as telling us something about the potential energy of 
these fields. In view of our current complete ignorance of what this
potential might be, we have tried, as much as possible, to 
understand generic properties, and we hope that the methods developed
here will be of further use in this respect. 
 
\bigskip{}
\noindent \textbf{Note added}
\bigskip

\noindent The interpretation of solutions of gravitational theories as
geodesics in a suitable metric space has a considerable history of
which we were mostly unaware at the time of writing this paper. The
geodesic interpretation of $k=0$ cosmologies that we have presented is
closely related to the Maupertuis-Jacobi principle of classical
mechanics. Relatively recent work on this topic includes
\cite{SmSo94,Gre95,CaGr96}. A particle mechanics formulation of the
geodesic interpretation described here, including our proposed
extension to $k\ne0$, has recently been found, and applied to a model
in which cosmological singularities correspond to horizons in the
augmented target space \cite{RuTo04}.    

\ack

PKT thanks Jaume Garriga for helpful discussions. The authors are
grateful to Lee Smolin for correspondence. MNRW gratefully
acknowledges financial support from the Gates Cambridge Trust.\\[15pt]


\begin{thebibliography}{99}
\bibitem{PeRa03}P.J.E. Peebles and B. Ratra. {\it The cosmological
  constant and dark energy.} Rev. Mod. Phys. {\bf 75} (2003) 559.
 
\bibitem{CC}
L. Cornalba and M. Costa. \textit{A new cosmological scenario in
  string theory.} Phys. Rev. {\bf D66} (2002) 066001. 

\bibitem{KKLT}
S. Kachru, R. Kallosh, A. Linde and S.P. Trivedi. \textit{De Sitter
  vacua in string theory.} Phys. Rev. {\bf D68} (2003) 046005. 

\bibitem{ToWo03}P.K. Townsend and
  M.N.R. Wohlfarth. \textit{Accelerating cosmologies from
    compactification.} 
Phys. Rev. Lett. \textbf{91} (2003) 061302.

\bibitem{Oht03a}N. Ohta. \textit{Accelerating cosmologies from
  S-branes}. Phys. Rev. Lett. \textbf{91} 
(2003) 061303.

\bibitem{Roy03}S. Roy. \textit{Accelerating cosmologies from M/string
  theory compactifications}. 
Phys. Lett. \textbf{B567} (2003) 322.

\bibitem{Woh03a}M.N.R. Wohlfarth. \textit{Accelerating cosmologies and
  a phase transition in 
M-theory}. Phys. Lett. \textbf{B563} (2003) 1.

\bibitem{Oht03b}N. Ohta. \textit{A study of accelerating cosmologies
  from superstring/M-theories}. 
Prog. Theor. Phys. \textbf{110} (2003) 269.

\bibitem{CHNW03}C.-M. Chen, P.-M. Ho, I.P. Neupane and
  J.E. Wang. \textit{A note on acceleration 
from product space compactification}. J. High Energy Phys. \textbf{07} (2003)
017.

\bibitem{CHNOW03}C.-M. Chen, P.-M. Ho, I.P. Neupane, N. Ohta and
  J.E. Wang. \textit{Hyperbolic 
space cosmologies}. J. High Energy Phys. \textbf{10} (2003) 058.

\bibitem{Woh03b}M.N.R. Wohlfarth. \textit{Inflationary cosmologies
  from compactification?} Phys. 
Rev. \textbf{D69} (2004) 066002. 

\bibitem{EmGa03}R. Emparan and J. Garriga. \textit{A note on
  accelerating cosmologies from compactifications 
and S-branes}. J. High Energy Phys. \textbf{05} (2003) 028.

\bibitem{Hal87}J.J. Halliwell. \textit{Scalar fields in cosmology with
  an exponential potential.} 
Phys. Lett. B \textbf{185} (1987) 341.

\bibitem{JMS04}L. J\"arv, T. Mohaupt and
 F. Saueressig. \textit{Quintessence cosmologies with 
a double exponential potential.} \texttt{hep-th/0403063}.


\bibitem{BB88}
A.B. Burd and J.D. Barrow. \textit{ Inflationary models with 
exponential potentials.}
Nucl. Phys. {\bf B308} (1988) 929. 

\bibitem{Tow03}P.K. Townsend. \textit{Cosmic acceleration and
  M-theory.} To appear in the proceedings 
of ICMP2003, Lisbon, Portugal, July 2003. \texttt{hep-th/0308149}.

\bibitem{Neu03}
I.P. Neupane. \textit{Accelerating cosmologies from exponential
  potentials.} \texttt{hep-th/0311071}. 

\bibitem{Vi03}
P.G. Vieira. \textit{Late-time cosmic dynamics from M-theory.} 
Class. Quantum Grav. {\bf 21} (2004) 2421.

\bibitem{Rus04}J.G. Russo. \textit{Exact solution of scalar-tensor
  cosmology with exponential 
potentials and transient acceleration.} \texttt{hep-th/0403010}.

\bibitem{CLW98}
E.J. Copeland, A.R. Liddle and D. Wands. 
\textit{Exponential potentials and cosmological 
scaling solutions.} Phys. Rev.  {\bf D57} (1998) 4686.

\bibitem{HeWa03}I.P.C. Heard and D. Wands. {\it Cosmology with positive
  and negative exponential potentials.} Class. Quantum Grav. {\bf 19}
  (2002) 5435.
  
\bibitem{Guo}Z.K. Guo, Y.-S. Piao and Y.-Z. Zhang. \textit{Cosmological
  scaling solutions and multiple exponential potentials.}
  Phys. Lett. {\bf B568} (2003) 1. 
  
\bibitem{BCGNR03}E. Bergshoeff, A. Collinucci, U. Gran, M. Nielsen and
  D. Roest. \textit{Transient  quintessence from group manifold 
reductions or how all roads lead to
  Rome.}  Class. Quantum Grav. {\bf 21} (2004) 1947.

\bibitem{TBF2002}S. Tsujikawa, R. Brandenberger and F. Finelli.
\textit{On the construction of nonsingular pre-big-bang and ekpyrotic
  cosmologies and the resulting density perturbations.}
Phys. Rev. \textbf{D66} (2002) 083513.

\bibitem{GiTo87}G.W. Gibbons and P.K. Townsend. 
\textit{Cosmological evolution of degenerate
vacua.} Nucl. Phys. \textbf{B282} (1987) 610.

\bibitem{DHN03}T. Damour, M. Henneaux and
  H. Nicolai. \textit{Cosmological billiards.} Class. Quantum
  Grav. \textbf{20} (2003) R145.

\bibitem{GHS87}
G.W. Gibbons, S.W. Hawking and J.M. Stewart. \textit{A natural measure
  on the set of all universes.}  
Nucl. Phys. {\bf B281} (1987) 736. 

\bibitem{ADS84}
I. Affleck, M. Dine and N. Seiberg. {\it Dynamical supersymmetry
  breaking in supersymmetric QCD.} Nucl. Phys. {\bf B241} (1984) 493.
 
\bibitem{DW67}

B.S. DeWitt. \textit{Quantum theory of gravity II: the 
manifestly covariant theory.}
Phys. Rev. {\bf 162} (1967) 1195. 

\bibitem{SmSo94}L. Smolin and C. Soo.
\textit{The Chern-Simons invariant as the natural time variable for
  classical and quantum cosmology.}
Nucl. Phys. \textbf{B449} (1995) 289.

\bibitem{Gre95}J. Greensite.
\textit{Field theory as free fall.}
Class. Quantum Grav. \textbf{13} (1996) 1339.

\bibitem{CaGr96}A. Carlini and J. Greensite.
\textit{The mass shell of the universe.}
Phys. Rev. \textbf{D55} (1997) 3514.

\bibitem{RuTo04}J.G. Russo and P.K. Townsend.
\textit{Cosmology as relativistic particle mechanics: from big crunch
  to big bang.}
\texttt{hep-th/0408220}.


\end{thebibliography}
\end{document}